\documentclass[submission,copyright,creativecommons]{eptcs}
\usepackage{amsmath}
\usepackage{amssymb}
\usepackage{tikz}

\providecommand{\event}{THedu'11} % Name of the event you are submitting to

\newtheorem{example}{Example}

\title{Formalization and Implementation of Algebraic Methods in Geometry}

\author{Filip Mari\' c, Ivan Petrovi\' c, Danijela Petrovi\' c, Predrag Jani\v ci\' c
\institute{Faculty of Mathematics,\\ University of Belgrade, Serbia}
\email{filip@matf.bg.ac.rs, mr00006@alas.matf.bg.ac.rs, danijela@matf.bg.ac.rs, janicic@matf.bg.ac.rs}
}
\def\titlerunning{Formalization and Implementation of Algebraic Methods in Geometry}
\def\authorrunning{Mari\' c et al.}

\begin{document}

\maketitle

\begin{abstract}
  We describe our ongoing project of formalization of algebraic
  methods for geometry theorem proving (Wu's method and the Gr\"obner
  bases method), their implementation and integration in educational
  tools. The project includes formal verification of the algebraic
  methods within Isabelle/HOL proof assistant and development of a
  new, open-source Java implementation of the algebraic methods.
  The project should fill-in some gaps still existing in this area
  (e.g., the lack of formal links between algebraic methods and
  synthetic geometry and the lack of self-contained implementations
  of algebraic methods suitable for integration with dynamic geometry
  tools) and should enable new applications of theorem proving in education.
\end{abstract}

% ***************************************************************************
\section{Introduction}
% ***************************************************************************

The field of automated deduction in geometry has been very successful
in the last several decades and a number of \emph{geometry automated
theorem provers (GATP)} have been developed. Most successful of these
are based on \emph{algebraic methods}, primarily Wu's method
\cite{wu1978} and the method of Gr\"obner bases
\cite{groebner_bases_and_applications}. The algebraic methods require
expressing geometric properties as polynomial equations in the
Cartesian plane, and then using algebraic techniques for dealing with
these equations. There are several implementations of these methods and
hundreds of complex geometry theorems have been proved automatically by
them \cite{chou}. However, despite these advances, there are still some
gaps in this area, preventing wider applications of the algebraic methods for
geometry, especially in education and in formal explorations of
geometry. Some of these gaps are:

\begin{itemize}
\setlength{\itemsep}{0pt}
\item There are no free, self-cointained, open source, and well-documented
  implementations of GATPs based on algebraic methods, suitable for
  integration with other tools (e.g., dynamic geometry tools).
\item There is no support for automated theorem proving in dynamic
  geometry tools most widely used on all levels of mathematical
  education (e.g., GeoGebra \cite{Hohenwarter04}) and this limits
  their applicability.
\item Only fragments of geometry and algebraic methods have been
  formalized within proof assistants (e.g., Isabelle, Coq), and there
  are still no formalized links between algebraic methods and synthetic
  geometry, giving formal correctness arguments for algebraic
  methods for geometry.
%\item There are still no standardized interchange formats for geometry
%  theorems (although there are initiatives for defining them) and
%  state-of-the art automated theorem provers use only their custom
%  input languages.
\end{itemize}

We aim at integrating algebraic methods in geometry with dynamic geometry
software (DGS), widely used in education, and in interactive proof assistants,
used for formalizing geometry. One of our goals is to address formalization
of algebraic methods and their integration into the interactive proof
assistant Isabelle/HOL. A major concern when using algebraic methods
is the lack of formally established connections between proofs
produced by these methods with classical synthetic proofs in geometry
(i.e., with Hilbert's or Tarski's geometry). We aim to make a strong,
formal link between algebraic methods and synthetic geometry by
formalizing both within Isabelle/HOL. Once the correctness of
algebraic methods is formally established within a proof-assistant,
these methods can be used as reflective tactics that help automating
significant portions of proofs in exploring geometry in a formal
setting. Our other goal is also to develop a corresponding Java implementations
of Wu's method and the Gr\"obner bases method conforming to open-source code
and documentation standards. These should directly support current
standardization initiatives for geometry formats (e.g., i2g, TGTP) and
should be suitable for integration into DGS. Both these goals should
enable new applications of (both automated and interactive) theorem provers
in education and in formalizing mathematics.

In this paper we report on the current status of our project. So far,
our Isabelle formalization includes formalization of the Cartesian
plane geometry, the translation of constructive geometry statements
into algebraic form, and connections between the algebraic form of a
statement and its interpretation in the Cartesian plane. Concerning
Java implementation, we have implemented Wu's method. Before presenting details
on this developments, we give a brief account on the related work and
on relevant background information. In all contexts, only plane (and
not space) geometry is considered.

{\em Overview of the paper.}
In Section \ref{sec:related} we present related work and state-of-the-art
results in the relevant areas; in Section \ref{sec:methods} we briefly
discuss algebraic methods and some common grounds for their formalization
and implementation; in Section \ref{sec:formalization} we present the
current status of our formalization of algebraic methods within Isabelle/HOL;
in Section \ref{sec:implementation} we present our Java implementation
of Wu's method; in Section \ref{sec:integrations} we discuss possible
integrations of our developments within a wider context; in Section
\ref{sec:education} we discuss potential applications in education,
and, in Section \ref{sec:conclusions} we draw some final conclusions.

% ***************************************************************************
\section{Related Work}
\label{sec:related}
% ***************************************************************************

\paragraph{Dynamic Geometry Tools.}
Dynamic geometry tools are mathematical software tools that allow
interactive work and visualization of geometric objects, by linking synthetic
geometry (most often --- Euclidean) with its standard models (e.g.,
Cartesian model). These tools are used in mathematical education,
but some of them are also used for producing digital
mathematical illustrations and animations. The common experience (not
scientifically supported, but with a number of positive empiric cases studies) is
that dynamic geometry tools significantly help students to acquire
knowledge about geometric objects. Some of the most popular dynamic
geometry tools are
{\it GeoGebra},\footnote{\url{http://www.geogebra.org}}
{\it Cinderella},\footnote{\url{http://www.cinderella.de}}
{\it Geometer's Sketchpad},\footnote{\url{http://www.keypress.com/sketchpad/}}
{\it Cabri},\footnote{\url{http://www.cabri.com}}
{\it GCLC},\footnote{\url{http://argo.matf.bg.ac.rs}}
{\it Eukleides}.\footnote{\url{http://www.eukleides.org}}
An overview of interactive geometry tools and their features can be found
on the Internet.\footnote{\url{http://en.wikipedia.org/wiki/Interactive_geometry_software}}

\paragraph{Automated Theorem Proving in Geometry.}
Automated theorem proving in geometry has a history more than fifty
years long \cite{handbook-ARG}. In the 1950s~Gelernter created a
theorem prover that could find solutions to a number of problems taken
from high-school textbooks in plane geometry \cite{gelertner}. The
biggest successes in automated theorem proving in geometry were
achieved (i.e., the most complex theorems were proved) by algebraic
provers based on Wu's method \cite{chou,wu1978} and Gr\"obner bases
method \cite{buchbergerPhd,Theorema06,Kapur86}. Instead of readable,
traditional geometry proofs, these methods produce only a yes/no
answer with a corresponding algebraic argument. Coordinate-free
methods, such as the area method \cite{Chou93,area-jar} and the full
angle method \cite{proof-in-geometry,Chou96a}, often produce readable
proofs, but for many conjectures these methods still deal with
extremely complex expressions involving certain geometry
quantities. An approach based on a deductive database and forward
chaining works over a suitably selected set of higher-order lemmas and
can prove complex geometry theorems, but still has a smaller scope
than algebraic provers \cite{proof-in-geometry}. Quaife used a
resolution theorem prover to prove theorems in Tarski's geometry
\cite{quaife89}. A theorem prover based on coherent-logic
\cite{CoherentLogic} can produce both readable and formal proofs of
geometry conjectures of a certain sort \cite{argoclp}.

\paragraph{Integration of Dynamic Geometry and Automated Theorem Proving.}
Just a few dynamic geometry systems have support for automated theorem
proving, typically based on Wu's method, the Gr\"obner bases method,
and the area method.
Geometry Expert\footnote{\url{http://www.mmrc.iss.ac.cn/gex/}} (GEX) is
a dynamic geometry tool focused on automated theorem proving and it
implements Wu's, the Gr\"obner bases, the vector, the full-angle, and
the area method \cite{gex1996}.
MMP/Geometer\footnote{\url{http://www.mmrc.iss.ac.cn/mmsoft/}}
is a new, Chinese, version of GEX \cite{MMP-geometer}.
It automates geometry diagram generation, geometry theorem proving,
and geometry theorem discovering.
Java Geometry Expert\footnote{\url{http://www.jgex.net/}} (JGEX) is a new,
Java version of GEX \cite{jgex}. It combines
dynamic geometry, automated geometry theorem proving, and, as its most
distinctive part, visual dynamic presentation of proofs. The systems from
the GEX family are publicly available, but only JGEX is open-source.
GEOTHER\footnote{\url{http://www-salsa.lip6.fr/~wang/GEOTHER}} a module of
Epsilon, is implemented in Maple, with drawing routines implemented in Java \cite{geother}.
Theorema\footnote{\url{http://www.theorema.org/}}
is a general mathematical tool, implemented in Mathematica, with a uniform framework
for computing, problem solving, and theorem proving \cite{Theorema06}. It has
some features of dynamic geometry systems and has support for automated theorem
proving in geometry \cite{robu_thesis}.
{\it Discover} is a system for automated proving and discovery in geometry, based
on two computer algebra systems, CoCoA and Mathematica \cite{Botana03}.
{\it Geometry Explorer} is a dynamic geometry tool proving support for the full-angle
method for automated theorem proving \cite{geometry-explorer}.
{\it GCLC}\footnote{\url{http://www.matf.bg.ac.rs/~janicic/gclc}}
is a geometry tool, implemented in C++, for visualizing geometry, for producing
mathematical illustration and for reasoning about geometry constructions. It is
based on a custom geometry language and provides support for three methods for
automated theorem proving \cite{gclc-jar}.
{\it GeoProof} is a dynamic geometry tool with built-in verified (within
Coq) geometry theorem prover based on the area method \cite{Narboux07}.

\paragraph{Formalizations of Geometry.}
There are a number of formalizations of fragments of various geometries
within proof assistants. Parts of Hilbert's seminal book {\em Foundations
of Geometry} have been formalized in Isabelle/Isar \cite{MeikleFleuriot2003}
and in Coq \cite{DehlingerDS00}. Within Coq, there are also formalizations
of von Plato's constructive geometry \cite{khan}, French high school geometry
\cite{guilhot2005}, Tarski's geometry \cite{Narboux2007}, ruler and compass
geometry \cite{duprat2002}, projective geometry \cite{MNS08}, etc. There
are efforts to integrate Gr\"obner bases solvers with proof-assistants
\cite{ChaiebW07,Pottier08}, but only the general part, not the one
dealing with geometry statements. Verified automated theorem proving in
geometry based on certificates has been developed in Coq for the
Gr\"obner bases method \cite{GregoirePT08} and for Wu's method
\cite{GenevauxNS11}.
In both these developments, certificates are generated using external provers.
Also, in both developments, there is no a proved link with synthetic geometry,
so --- substantially --- only algebraic conjectures are considered.
We are not aware of a full algebraic method for geometry theorem proving
formalized within a theorem prover. We are also not aware of a formalized
relationship between algebraic methods and synthetic geometry --- i.e.,
of formalized claim that if some geometry conjecture is proved by an
algebraic method, then it is theorem of synthetic geometry.

% ***************************************************************************
\section{Algebraic Methods in Geometry}
\label{sec:methods}
% ***************************************************************************

Algebraic methods in geometry are well described in literature
\cite{proof-in-geometry,chou,WangBook,wu1978} and many of its aspects
are already considered to be folklore. Still, some issues deserve an
attention, especially in the context of interactive theorem
proving. In this section we give a brief account on algebraic methods
in geometry, in order to make the paper self-contained, but also to
stress some issues relevant for formalizing the link between synthetic
geometry and the algebraic methods.

% ----------------------------------------------------------------------------
\subsection{Synthetic Euclidean Geometry and Analytic (Cartesian) Geometry }
% ----------------------------------------------------------------------------

Synthetic Euclidean geometry is geometry based on a, typically small,
set of primitive notions and axioms. There is a number of variants of
axiom systems for Euclidean geometry and the most influential and
important ones are Euclid's system (from his seminal ``Elements'')
and its modern reincarnations \cite{AvigadElements}, Hilbert's system
\cite{hilbert-geometrie}, and Tarski's system \cite{tarski83}.
Tarski's geometry is built in first-order logic and is less powerful
than Hilbert's one (built in higher-order logic).

In analytic (or Cartesian) geometry, points are represented as pairs
of real numbers (\emph{Cartesian coordinates}) and lines and curves as
algebraic equations. All axioms of Euclidean geometry are
valid (using the standard, natural interpretation) in Cartesian plane,
so this plane is a model of Euclidean geometry and each Euclidean
theorem is valid in Cartesian plane. It can be proved that the
opposite also holds: each statement valid in Cartesian plane is a
theorem of (any reasonably built) synthetical geometry.

% -----------------------------------------------------------------------
\subsection{Algebraization of Geometry Statements}
\label{sec:Algebraization}
% -----------------------------------------------------------------------

Algebraic methods, used as methods for automated theorem proving in
geometry for theorems of constructive type (i.e., conjectures about
geometric objects obtained by geometric constructions), introduce
(symbolic) coordinates for geometric objects involved (points, and
possibly lines), express geometric constructions and statements as
algebraic (multivariate polynomial) equations involving introduced
coordinates and then use algebraic means to prove that the statement
follows from the construction.

The standard algebrization procedure introduces fresh symbolic
variables for point coordinates and introduces (polynomial) equations
that characterize every construction step and the statement to be
proved. Although for lines involved in the construction unknown
coefficients could be introduced, the standard procedure avoids that
and uses only points (while lines are specified only implicitly). Each
construction starts from a set of \emph{free points} and introduces
\emph{dependent points} along the way. In some cases, dependent points
are chosen with a degree of freedom (e.g., choosing a random point on
line). Each point gets a pair of coordinates represented by symbolic
variables. Free variables are usually denoted by $u_i$
($i=0,1,2,\ldots$), while the dependent ones are denoted by $x_i$
($i=0,1,2,\ldots$). If a point is free, both its coordinates will be
free variables. If a point is, dependent, but with one degree of
freedom, one coordinate will be a free, while the another one will be
a dependent variable.\footnote{However, choosing which one should be
  free and which one dependent is not trivial and requires special
  attention. For example, given that $A$ is a point on a line $l$, one
  of its coordinates could be free while the other is dependent and
  calculated from the line constraints. However, if $l$ is parallel to
  the $x$-axis, then the $y$ coordinate of the point $A$ cannot be
  free. Also, if $l$ is parallel to the $y$-axis then the $x$
  coordinate of the point $A$ cannot be free.} If a point is
dependent both its coordinates will be dependent ones.

Geometry constraints over points can be formulated as algebraic
constraints over the point coordinates (i.e., as polynomial equations
over the introduced symbolic variables). For example, assume that
symbolic coordinates of the point $A$ are $(x_a, y_a)$, the point $B$
are $(x_b, y_b)$, and the point $C$ are $(x_c, y_c)$. The fact that
$A$ is the midpoint of the segment $BC$ corresponds to the algebraic
conditions $2 x_a = x_b + x_c$ and $2 y_a = y_b + y_c$. The
fact that $A$, $B$, and $C$ are collinear corresponds to the algebraic
condition $(x_a-x_b)(y_b-y_c) = (y_a-y_b)(x_b-x_c)$. Similar
connections are formulated for other basic geometry relationships
(parallel lines, perpendicular lines, segment bisectors, etc.).

\begin{example}
\label{ex:midsegment}
  Let $ABC$ be a triangle, and let $B_1$ be the midpoint of the edge
  $AC$ and $C_1$ be the midpoint of the edge $AB$. Then, the
  midsegment $B_1C_1$ is parallel to $BC$.

\begin{figure}[ht!]
  \centering
  \input{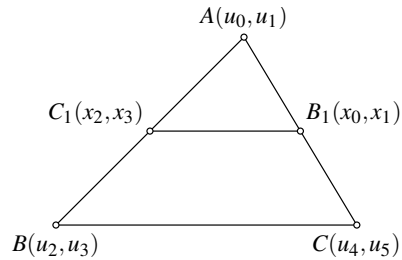}
  \caption{Midsegment theorem}
  \label{fig:midsegment}
\end{figure}

In this example, $A$, $B$, and $C$ are free points so they are
introduced symbolic variables $A(u_0, u_1)$, $B(u_2, u_3)$, and
$C(u_4, u_5)$. Points $B_1$ and $C_1$ are dependent so they are
introduced symbolic variables $B_1(x_0, x_1)$ and $C_1(x_2, x_3)$.
Since $B_1$ is the midpoint of $AC$, it holds that $2 x_0 = u_0 +
u_4$ and $2 x_1 = u_1 + u_5$. Since $C_1$ is the midpoint of
$AB$, it holds that $2 x_2 = u_0 + u_2$ and $2 x_3 = u_1 +
u_3$. These four equations come from the description of the
construction, i.e., from the premises of the conjecture. In order to
show that $B_1C_1$ is parallel to $BC$, it suffices to show that
$(x_2-x_0)(u_5-u_3)\,=\,(x_3-x_1)(u_4-u_2)$
holds. This equation corresponds to the conclusion of the
conjecture. So, the geometric problem is reduced to showing that every
n-tuple satisfying the first four equations (stemming from the
construction) also satisfies the last equation (stemming from the
conclusion), i.e., to show that
$$
\begin{array}{ll}
\forall u_0\;u_1\;u_2\;u_3\;u_4\;u_5\;x_0\;x_1\;x_2\;x_3 \in \mathbb{R}.& 2 x_0 = u_0 + u_4 \,\wedge\, 2 x_1 = u_1 + u_5 \,\wedge\,
2 x_2 = u_0 + u_2 \,\wedge\, 2 x_3 = u_1 + u_3\\
&  \implies (x_2-x_0)(u_5-u_3)\,=\,(x_3-x_1)(u_4-u_2).
\end{array}
$$
\end{example}

Note that in the above example, the condition that $ABC$ is a triangle
is not translated into conditions that $A$, $B$, $C$ are pairwise
different. Also, the condition that $B_1C_1$ is parallel to $BC$
is represented by equation $(x_2-x_0)(u_5-u_3)\,=\,(x_3-x_1)(u_4-u_2)$.
However, this algebraic equation is actually equivalent to
the following, weaker condition: $B \equiv C$ or $B_1 \equiv C_1$ or
$B_1C_1$ is parallel to $BC$. Therefore, when translated back in
geometry terms, the conjecture that is to be proved by an algebraic
method is:

{\em Let $B_1$ be the midpoint of the segment $AC$ and $C_1$ be the midpoint of
the segment $AB$. Then, the segment $B_1C_1$ is parallel to $BC$ or $B$ is
identical to $C$ or $B_1$ is identical to $C_1$.}

Since, $B_1 \not\equiv C_1$ follows from $B \not\equiv C$, the above
conjecture is equivalent with

{\em Let $B$ and $C$ be two distinct points, let $B_1$ be the
  midpoint of the segment $AC$ and $C_1$ be the midpoint of the
  segment $AB$. Then, the segment $B_1C_1$ is parallel to $BC$.}

This example shows that translating a conjecture from geometry terms to algebraic
terms and vice versa involves dealing with important details. In most systems,
a hypothesis of the form $AB \| CD$ is typically represented by the equation
of the form $(x_b-x_a)(y_d-y_c)\,=\,(x_d-x_c)(y_b-y_a)$ and moreover, this
equation is used as a definition for $AB \| CD$. However, this approach,
unfortunately, breaks the link with synthetic geometry.

It can be shown that most geometry properties are invariant under
isometric transformations \cite{GenevauxNS11,Harrison-handbook}. If
$P_1$ and $P_2$ are two free points, there always exists an isometry
(more precisely, a composition of a translation and a rotation) that
maps $P_1$ to the point $(0,0)$ (i.e., the origin of the Cartesian
plane) and $P_2$ to a point on the $x$-axis. Therefore, without loss
of generality, it can be assumed that one free point has coordinates
$(0, 0)$, while another one has coordinates $(u_0,0)$ or $(0,u_0)$
(although both these choices are correct, in some cases the choice
could affect efficiency, or, if the partial Wu's method is used,
the choice could even determine
whether it will be possible to do the proof). This assumption can
significantly reduce the amount of work needed by the algebraic
methods. In addition, there are heuristics (aimed at improvement of
efficiency) for choosing among the free points which two will get
these distinguished coordinates.

\begin{example}
Without loss of generality in the conjecture from Example \ref{ex:midsegment},
the points $B$ and $C$ can be assigned coordinates $B(0,0)$, and $C(u_4,0)$.
Therefore, the algebraic conjecture to be proved is as follows:

\vspace{-0.5cm}
$$
\begin{array}{l}
\forall u_0\;u_1\;u_4\;x_0\;x_1\;x_2\;x_3 \in \mathbb{R}. \\
2 x_0 = u_0 + u_4 \,\wedge\, 2 x_1 = u_1 \,\wedge\,
2 x_2 = u_0 \,\wedge\, 2 x_3 = u_1 \implies (x_2-x_0) \cdot 0\,=\,(x_3-x_1) u_4.
\end{array}
$$

\noindent
which is trivially valid (since $x_3-x_1=0$ follows from
$2 x_1 = u_1$ and $2 x_3 = u_1$).
\end{example}

%-----------------------------------------------------------------
\subsection{Algebraic Algorithms}
%-----------------------------------------------------------------

Once the geometric theorem has been algebrized, algebraic theorem
proving methods themselves can be applied. Algebraic theorem provers
use specific algorithms over polynomial systems (each polynomial
equation of the form $p_1(v_1, \ldots, v_n) = p_2(v_1, \ldots, v_n)$
is transformed to $p_1(v_1, \ldots, v_n) - p_2(v_1, \ldots, v_n) = 0$,
i.e., to the form $p(v_1, \ldots, v_n) = 0$). If $f_1$, \ldots, $f_k$
are polynomials obtained from the construction, and $g_1$, \ldots,
$g_l$ are polynomials obtained from the statement, then the conjecture
is reduced to checking if for each $g_i$ it holds that
$$\forall v_1, \ldots, v_n \in \mathbb{R}.\ \bigwedge_{i=1}^k f_i(v_1, \ldots, v_n) = 0 \implies g_i(v_1\ldots v_n) = 0.$$

As Tarski noted, this could be decided by a quantifier elimination
procedure for the reals. In practice, it is hard to prove nontrivial
geometric properties in this fashion, because even sophisticated
algorithms for real quantifier elimination are relatively inefficient.
Therefore, another approach is taken.
The main insight, given by Wu Wen-ts\"un in 1978, is that remarkably
many geometrical theorems, when formulated as universal algebraic
statements in terms of coordinates, are also true for all complex
values of the ``coordinates''. So, instead of checking polynomials over
reals, the field of complex numbers is used and the following conjecture
is considered\footnote{Of course, this leads to the incompleteness (with
  respect to geometry) of the methods since, in some cases, the
  condition holds for $\mathbb{R}$, while the methods fails to detect
  that due to some counterexamples from $\mathbb{C}$.}:

\begin{equation}
\label{eq:alg}
\forall v_1, \ldots, v_n \in \mathbb{C}.\ \bigwedge_{i=1}^k f_i(v_1, \ldots, v_n) = 0 \implies g_i(v_1\ldots v_n) = 0.
\end{equation}

This is true when $g$ belongs to the radical of the ideal $I = \langle f_1,
\ldots, f_k \rangle$, generated by the polynomials $f_i$, i.e., when there
exists an integer $r$ and polynomials $h_1, \ldots, h_k$ such that
$g_i^r = \Sigma_{i=1}^kh_if_i$. Hilbert's Nullstellensatz
theorem states that if the field is algebraically closed (as
$\mathbb{C}$ is), then the converse is also true i.e., such
decomposition can always be made.

The two most significant algebraic methods use a kind of Euclidean division to
check the validity of a conjecture of the form \ref{eq:alg}.
Buchberger's method consists in transforming the generating set
into a {\em Gr\"obner basis}, in which a division algorithm can
be efficiently used, while in the Wu's method
a \emph{pseudo-division} is used which closely mimics Euclidean
division.

The main operation over polynomials in Wu's method is pseudo division
which, when applied to two polynomials $p(v_1, \ldots, v_n)$ and
$q(v_1, \ldots, v_n)$ produces the decomposition
$$c^rp = tq + r,$$
where $c(v_1, \ldots, v_{n-1})$ is the leading coefficient of $q$ in
the variable $v_n$, $r$ is the number of non-zero coefficients of $p$,
$t(v_1, \ldots, v_n)$ is the pseudo-quotient, $r(v_1, \ldots, v_n)$ is
the pseudo-remainder, and the degree $v_n$ in $r$ is smaller then in
$q$. Since $r = c^rp - tq$, it is clear that $r$ belongs to the ideal
generated by $p$ and $q$.

The first step of Wu's method (the simple form \cite{chou}) uses the pseudo-division operation to
transform the construction polynomial system ($\bigwedge_{i=1}^k f_i$)
to triangular form, i.e., to a system of equations where each
successive equation introduces exactly one dependent variable. After
that, the final reminder is calculated by pseudo dividing polynomial
for statement ($g_i$) by each polynomial from triangular system.

Summarizing, Wu's method, in its simplest form, allows to compute
some polynomials $c$, $h_1, \ldots, h_k$ and $r$ such that
$$cg_i = \sum_{i=1}^kh_if_i + r.$$ If the final remainder $r$ is equal to zero,
then the conjecture is considered to be proved. This simple method of
Wu is not complete (in algebraic sense). A more complex and complete
version of the method uses ascending chains which are considered in
the Ritt-Wu principle.

%--------------------------------------------------------------------
\subsection{Non-degeneracy Conditions}
\label{subsec:ndg}
%--------------------------------------------------------------------

An important feature of algebraic theorem proving methods for
geometry is that they automatically provide conditions --- {\em
  non-degeneracy} or {\em NDG conditions} --- under which the given
statement holds. For example, once the decomposition $cg_i =
\sum_{i=1}^kh_if_i$ in the simple Wu's method is obtained, if $c(v_1,
\ldots, v_n) \neq 0$ it holds that $\bigwedge_{i=1}^k f_i(v_1, \ldots,
v_n) = 0 \implies g_i(v_1, \ldots, v_n) = 0$. However, if $c(v_1,
\ldots, v_n) = 0$, then this conclusion cannot be made, so the
conjecture is true only for those n-tuples for which a degeneracy
(given by $c$) is not present. NDG conditions generated by algebraic
methods are given in algebraic terms, but in the context of geometry
theorem proving, it is important to have their geometrical
interpretation as well. With NDG conditions given in the form of
geometrical assumptions, one gets the final geometry theorem, avoiding
some {\em degenerate cases} that make the original conjecture
invalid. However, it still might be the case that the statement is
valid under some weaker assumptions. In addition, a presence of NDG
conditions doesn't necessarily mean that if they are not satisfied the
conjecture is invalid.

In Wu's method, NDG conditions are obtained from the triangular system as
leading coefficients of variables introduced in each polynomial.
These coefficients are actually $x$-polynomials over other variables
(introduced in triangular system in previous equations). In the
Gr\"obner basis method, NDG conditions are $u$-polynomials that are
denominators of $u$-fractions which are coefficients of
$x$-polynomials obtained during process of reduction to Gr\"obner
basis.

\begin{example}
  In a triangle $ABC$, let $h_a$, $h_b$, and $h_c$ be altitudes that
  correspond to the vertices $A$, $B$, and $C$ and let $H$ be the
  intersection of $h_a$ and $h_b$. Then, $H$ belongs to $h_c$.  The
  conjectures can be proved by algebraic methods in the following way.

\begin{figure}[ht]
  \centering
  \input{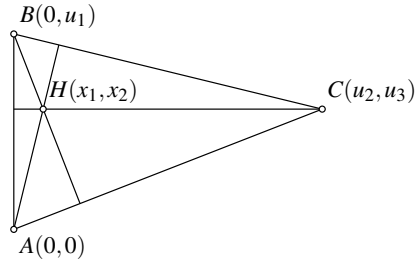}
  \caption{Orthocenter theorem}
  \label{fig:orthocentre}
\end{figure}

Let the points $A$, $B$, $C$, and $H$ be assigned coordinates
$(0,0)$, $(u_{1},0)$, $(u_{2},u_{3})$, $(x_{1}, x_{2})$.  The
condition that $AH$ is perpendicular with the line $BC$ is represented
by: $(x_1-0)(u_2-0) + (x_2-0)(u_3-u_1) = 0$ i.e., $(u_3-u_1)x_2 +
u_2x_1=0$. The condition that $BH$ is perpendicular with the line $AC$
is represented by: $(x_1-0)(u_2-0)+(x_2-u_1)(u_3-0) = 0$, i.e.,
$u_3x_2 + u_2x_1 - u_3u_1=0$.  The statement of the conjecture
corresponds to the condition $x_2 - u_3=0$. After the triangulation,
the polynomial system becomes:

$-u_2u_1x_1 + (-u_3^2u_1+u_3u_1^2) = 0$,

$(u_3-u_1)x_2 + u_2x_1 = 0$

After pseudo-division of $x_2 - u_3=0$ with the above polynomials, the
pseudo remainder is $0$, which prove the conjecture, but only under
conditions $-u_2u_1 \neq 0$ and $u_3-u_1 \neq 0$, i.e., $u_1 \neq 0$,
$u_2 \neq 0$, $u_1 \neq u_3$, corresponding to geometry constraints
that $A \not\equiv B$, $A$, $B$ and $C$ are not collinear, and $BC
\not\perp AB$.

It is interesting to notice that the computed NDGs can be weakened:
the last two conditions can be omitted. Indeed, for all values that
make the construction polynomials equal to zero, the statement polynomial is also
equal to zero. Moreover, the geometry counterpart is also valid under
these weakened NDG conditions, assuming that $XY$ denotes \emph{some}
line containing both $X$ and $Y$ (which is relevant in some
degenerated cases).
\end{example}

The subtle questions about NDG conditions in algebraic and geometry
terms \cite{WangBook} can be rigorously analyzed within formalization
of algebraic methods.

As said, NDG conditions give {\em sufficient} (i.e., conditions under
which calculations applied by Wu's method are allowed -- they prevent
pseudo division with the zero polynomial) and not {\em necessary}
conditions for the (extended) conjecture to be valid. We illustrate
this by the following example (Simpson's theorem).

\begin{example}
Let $ABC$ be a triangle and $D$ is arbitrarily chosen point on
circumscribed circle of this triangle. If $K$, $L$ and $M$ are
respectively foot points of perpendicular lines through $D$ to edges
$AB$, $BC$ and $CA$ of triangle $ABC$, then these three points $K$,
$L$ and $M$ are collinear.

Let us assume that the following symbolic coordinates are associated
to the introduced points: $A(0, 0)$, $B(0, u_{1})$, $C(u_{2}, u_{3})$,
$D(u_{4}, x_{1})$, $M(0, x_{1})$, $N(x_{2}, x_{3})$, $P(x_{4}, x_{5})$.
The construction corresponds to the following system of polynomial equations:

\begin{eqnarray*}
 & & u_{2}x_{1}^{2}-u_{2}u_{1}x_{1} +
(u_{4}^{2}u_{2}-u_{4}u_{3}^{2}+u_{4}u_{3}u_{1}-u_{4}u_{2}^{2})  = 0 \\
 & & (u_{3}-u_{1})x_{3} + u_{2}x_{2} + (-u_{3}+u_{1})x_{1}-u_{4}u_{2} = 0 \\
 & & u_{2}x_{3} + (-u_{3}+u_{1})x_{2}-u_{2}u_{1} = 0 \\
 & & u_{3}x_{5} + u_{2}x_{4}-u_{3}x_{1}-u_{4}u_{2} = 0 \\
 & & u_{2}x_{5}-u_{3}x_{4} = 0 \\
\end{eqnarray*}

\noindent
and the polynomial equation for theorem statement is $x_{5}x_{2}-x_{4}x_{3} + x_{4}x_{1}-x_{2}x_{1} = 0.$

Following Wu's method, the following NDGs can be extracted: $u_{1} \neq 0$, $u_{2} \neq 0$, $u_{3} \neq 0$,
$u_{3}^{2}-2u_{3}u_{1}+u_{2}^{2}+u_{1}^{2} \neq 0$, $u_3\neq u_1$, $u_{3}^{2}+u_{2}^{2} \neq 0$.

The NDG condition $u_{3}^{2}-2u_{3}u_{1}+u_{2}^{2}+u_{1}^{2} \neq 0$
can be interpreted as the following geometry condition: {\it Line
  through points $C$ and $B$ is not perpendicular to line through
  points $B$ and $A$}, or in other words, the triangle $ABC$ is not
right triangle with right angle at vertex $B$. However, this is not a
degenerate geometry case, but rather a condition required by the
application of Wu's method. Indeed, the theorem holds in this special
case as well. It can be proved if the point $C$ is constructed on the
line through $B$ perpendicular to line $AB$. Then, the points are
instantiated as follows: $A(0, 0)$, $B(0, u_{1})$, $C(u_{2}, u_{1})$,
$D(u_{3}, x_{1})$, $M(0, x_{1})$, $N(u_{3}, u_{1})$, $P(x_{2}, x_{3})$
and the problem is transformed to following polynomial system:

\begin{eqnarray*}
 & & x_{1}^{2}-u_{1}x_{1} + (u_{3}^{2}-u_{3}u_{2})  = 0 \\
 & & u_{1}x_{3} + u_{2}x_{2}-u_{1}x_{1}-u_{3}u_{2} = 0 \\
 & & u_{2}x_{3}-u_{1}x_{2} = 0 \\
\end{eqnarray*}

\noindent
and with polynomial for statement $u_{3}x_{3} +
x_{2}x_{1}-u_{1}x_{2}-u_{3}x_{1} = 0 .$

The theorem is proved in this case, with the following two NDG conditions:
$u_{2}^{2}+u_{1}^{2} \neq 0$, $u_{1} \neq 0$.

\noindent
which is equivalent to trivial NDG conditions $u_{1} \neq 0$, $u_{2} \neq 0$.
\end{example}

% ***************************************************************************
\section{Formalization of Algebraic Methods within Isabelle/HOL}
\label{sec:formalization}
% ***************************************************************************

An important part of our project is to develop a formalization of
algebraic methods in Isabelle/HOL.\footnote{The main authors are Filip
  Mari\' c and Danijela Petrovi\' c. The current proof documents are
  available at \url{http://argo.matf.bg.ac.rs}.} Our goals are:

\begin{itemize}
\setlength{\itemsep}{0pt}
\item to identify and separate relevant geometry and algebra concepts
  and formalize them;
\item to build formally verified geometry axiomatizations and models;
\item to identify key shared features of different algebraic methods;
\item to implement algebraic GATPs that are formally verified, yet
  efficient enough to handle non-trivial geometric statements;
\item to integrate algebraic GATPs to Isabelle/HOL and facilitate
  their use in education and in formal explorations in geometry.
\end{itemize}

In this section we present first steps and some choices made towards
this formalization. Although our final goals require establishing a
connection between algebraic methods and synthetic geometries, at the
current stage only the connection between algebraic methods and
analytic geometry is established. However, many early choices in the
formalization are made having in mind connections with synthetic
geometries that will be analyzed in the further stages of our work,
and some layers of abstraction are introduced to our formalization
having in mind these goals.

\subsection{Representation of Geometry Conjectures}

Geometry conjectures (of constructive type) are represented by terms
over an abstract syntax. To emphasize the constructive nature, along
standard geometrical relations (e.g., collinearity, parallelness,
perpendicularity), the syntax supports functions that construct new
objects starting from the given ones (e.g., construct the midpoint of
a segment, construct the intersection of two lines, construct a line
trough a given point perpendicular to a given line). This makes the
system close to the dynamic geometry tools like GCLC \cite{gclc-jar}
and GeoGebra \cite{Hohenwarter04}
that naturally support this kind of constructions.

Within Isabelle/HOL, the abstract syntax is defined by mutually recursive
datatypes \verb|point_term| (with constructors such as
\texttt{MkPoint}, \texttt{MkMidpoint}, \texttt{MkIntersection}, etc.)
corresponding to point constructions, \verb|line_term| (with
constructors such as \texttt{MkLine}, \texttt{MkPerp},
\texttt{MkParallel}, etc.) corresponding to line constructions, and
\verb|statement_term| (with constructors like \texttt{Collinear},
\texttt{Congruent}, \texttt{Parallel}, \texttt{Perp}, etc.)
corresponding to geometrical relations. In this syntax, for example,
the statement that three perpendicular bisectors of edges of a
triangle $ABC$ meet in a single point can be represented by the
following HOL term (the points \verb|A|, \verb|B|, \verb|C| are free
points):

\begin{small}
{\tt
\begin{tabbing}
let \=\kill\\
let A = MkPoint 1; B = MkPoint 2; C = MkPoint 3;\\
  \> C1 = MkMidpoint A B; A1 = MkMidpoint B C; B1 = MkMidpoint C A;\\
  \>  O = MkIntersection (MkPerp (MKLine A B) C1) (MkPerp (MkLine B C) A1) in\\
  \>  Incident O (MkPerp (MkLine C A) B1)
\end{tabbing}
}
\end{small}

\subsection{Interpreting Terms in Geometry Models}

Syntactic terms that represent geometry conjectures can be interpreted
in different geometries (e.g., Cartesian geometry, Hilbert's geometry,
Tarski's geometry). For convenience, Isabelle's \emph{locales}
infrastructure is used to avoid repeating definitions. A locale
\verb|AbstractGeometry| is defined that contains
primitive\footnote{Deciding whether some notion is primitive or
  derived depends on the underlying geometry. We chose to use a very
  high level abstraction (with more primitive notions) to separate the
  correctness concerns of algebraic methods with respect to analytic
  geometry from concerns about the internal connections of geometric
  notions. Separation of primitive and derived notions will be handled
  within formalisation of each specific geometry.} relations needed to
interpret a geometric statement (e.g., {\em incident}, {\em between},
{\em congruent}, {\em perpendicular}, {\em parallel}) and postulates
their properties.  Derived concepts are be defined only within this
locale. For example, the (derived) notion of collinearity reduces to
(primitive) notion of incidency --- three points are said to be
collinear iff there is a line that is incident with all three
points. Various geometries interpret this locale and (abstract)
definitions of derived notions transfer to these geometries.

Semantics of a term (in abstract geometry) is given by functions
\verb|point_interp|, \verb|line_interp| and \verb|statement_interp|,
that take a \verb|point_term|, \verb|line_term| or a
\verb|statement_term| and return a (abstract) point, a (abstract) line
or a Boolean value, respectively. Since abstract interpretation of a
term is uniquely determined only if the free points are fixed, all these
functions take an additional argument --- a function mapping free
point indices to points. Since in some degenerate cases it is not
possible to assign a (unique) geometry object to a term (e.g., when
interpreting a term that constructs intersection of two parallel
lines) construction functions are partial --- the interpretation
assigns Isabelle/HOL's \texttt{option} type values to terms i.e., the
term can be interpreted either by the value \texttt{None} (when no
object could be used as interpretation of the term) or \texttt{Some o}
(when the object \texttt{o} is the interpretation of the term).

Statements are interpreted by the primitive relations of the abstract
geometry, while constructions are reduced to primitive relations using
the Hilbert's choice operator ($\varepsilon$ or \verb|SOME| in
Isabelle)\footnote{Note that this decision enables interpreting
  constructions that do not have unique interpretations (e.g., it is
  possible to construct an intersection of two identical lines,
  yielding an arbitrary point on those lines), while in the literature
  these kinds of constructions are sometimes considered
  degenerate.}. For example:

\begin{small}
{\tt
\begin{tabbing}
xxx\=xxx\=\kill
statement\_interp (Incident p l) fp =\\
\> let pi = point\_interp p fp; li = line\_interp l fp in\\
\> if (pi $\neq$ None $\wedge$ pi $\neq$ None) then\\
\>\> incident (the pi) (the li)\\
\>  else\\
\>\>  None\\
\\
point\_interp (MkIntersection l1 l2) fp =\\
\> let l1i = line\_interp l1 fp; l2i = line\_interp l2 fp in\\
\> if (l1i $\neq$ None $\wedge$ l2i $\neq$ None $\wedge$ $\exists$ P. incident P (the l1i) $\wedge$ incident P (the l2i)) then\\
\>\>  Some (SOME P. incident P (the l1i) $\wedge$ incident P (the l2i))\\
\> else\\
\>\>  None
\end{tabbing}
}
\end{small}

A statement (represented by a term) is valid in a (abstract) geometry
if all its interpretations (i.e., interpretations for all choices of
free points) that are not trivially degenerate are true.

\begin{small}
{\tt
\begin{tabbing}
\hspace{2ex}\=\hspace{10ex}\=\kill
definition (in AbstractGeometry) valid :: "statement\_term $\Rightarrow$ bool" where\\
\>"valid stmt = $\forall$ fp. let si = statement\_interp stmt fp in\\
\>\>                         si $\neq$ None $\longrightarrow$ si = Some True"
\end{tabbing}
}
\end{small}

All the previous notions lift to concrete geometries (e.g., Cartesian
geometry, Hilbert's geometry, Tarski's geometry) once it is shown that
these are interpretations of the \verb|AbstractGeometry| locale.

\subsection{Formalizing Cartesian Geometry}

Cartesian geometry has been defined within Isabelle/HOL. Points
are defined as pairs of real numbers. Lines are determined by their
equations of the form $Ax + By + C = 0$, so a line is defined as an
equivalence class of triplets $(A, B, C)$ where $A \neq 0 \vee B \neq
0$ and triplets are equivalent iff they are proportional. This kind of
reasoning has been facilitated by support for quotient types and
quotient definitions that has been recently introduced to
Isabelle/HOL \cite{isabelle-quotient}.

Once the types of points and lines are introduced, they are used to
interpret the \verb|AbstractGeometry| locale by introducing primitive
geometric predicates and proving their properties (basically
axiom-level statements). For example, the predicate {\em incident} is
defined by a quotient definition, stating that a point $(x, y)$ is
incident to a line determined by the coefficients $(A, B, C)$ iff $A x +
B y + c = 0$, and for example, it is proved that there is a
unique line incident to two different points.

\subsection{Algebrizing Geometry Conjecture Terms}

We have implemented a recursive algebraization procedure
(outlined in Section \ref{sec:Algebraization}) within
Isabelle/HOL. It keeps track of the current state containing symbolic
coordinates of points introduced during the construction (these are
subterms of the original statement term) and polynomials (both for
construction and the statement) obtained so far.

The function \verb|algebrize| takes a statement term and returns sets
of construction and statement polynomials obtained by our
algebrization procedure, starting from an empty initial state. The
procedure descends the term recursively and adds specific polynomials
for each construction step or for the statement. For example, when
algebrizing a statement of the form \verb|Colinear A B C|, the set of
statement polynomials includes the polynomial
$(x_a-x_b)(y_b-y_c)-(y_a-y_b)(x_b-x_c)$, where $(x_a, y_a)$,
$(x_b, y_b)$, and $(x_c, y_c)$ are symbolic coordinates introduced
during algebrization of the constructions for point terms \verb|A|,
\verb|B|, and \verb|C| (during this, the set of construction
polynomials was appropriately extended so that it contains definitions
of these symbolic coordinates in term of symbolic coordinates of free
points).

The central theorem that we have formally proved is that, if all
statement polynomials are equal to zero whenever all construction
polynomials are (i.e., if all statement polynomials belong to the
radical of ideal generated by the construction polynomials), then the
statement is valid in analytic geometry.

\begin{small}
{\tt
\begin{tabbing}
\hspace{2ex}\=\hspace{10ex}\=\kill
theorem "let (cp, sp) = algebrize term in\\
\>($\forall$ ass.\>(($\forall$ p $\in$ cp.\ eval\_poly ass p = 0) $\longrightarrow$ p $\in$ sp.\ eval\_poly ass p = 0)) $\longrightarrow$\\
\>\>     AnalyticGeometry.valid s)"
\end{tabbing}
}
\end{small}

\subsection{Algebraic Algorithms and Non-degeneracy Conditions}

Currently, for checking polynomial ideal membership problem we use the
Gr\"obner bases solver already integrated into Isabelle/HOL. It is a
general purpose solver, not aimed at geometry, so it is not capable of
finding the required NDG conditions.  Therefore, the user can either
specify NDGs or use the most general NDGs that can be inferred from
the construction description.

\subsection{Further Work: Connections with Synthetic Euclidean Geometries}

Goals of the presented project also include formalizing the
connections between algebraic methods and synthetic axiomatizations
(established via Cartesian plane). We will consider only full
Euclidean geometry (and not affine geometry). The first step would be
to formalize Tarski's elementary geometry within Isabelle/HOL. Then,
it would be necessary to prove that every valid statement in analytic
geometry also hold in all models of Tarski's geometry. The central
(most demanding) part would then be to formally prove the deductive
completeness of Tarski's geometry (by formally establishing the
connection with real-closed fields and formally showing that they
allow quantifier-elimination). The next step would be to analyze the
connections with Hilbert's geometry by showing that all proofs in
Tarski's geometry can be transferred to Hilbert's geometry (since all
Tarski's axioms can be proved within Hilbert's geometry). Finally, if
an algebraic method is proved to be correct (in purely algebraic
terms), then it would follow that it is correct when used for geometry
theorem proving.

% ***************************************************************************
\section{Implementation of Algebraic Methods in Java}
\label{sec:implementation}
% ***************************************************************************

In this section we briefly describe our open-source, documented OpenGeoProver
software implemented in Java,\footnote{The main author of the
  implementation is Ivan Petrovi\' c. The source code is available at
  \url{http://argo.matf.bg.ac.rs}.} aimed at providing self-contained
support for algebraic theorem proving methods for geometry and, also,
for Wu's method and the Gr\"obner bases method themselves. OpenGeoProver
is an improved reimplementation of the algebraic methods implemented
in C++ within the {\it GCLC} tool \cite{Predovic08}. Currently, only
Wu's method is supported, but we are planning to implement the
Gr\"obner bases method as well. Currently, there is no implementation
of B-reduction, the main operation in Gr\"obner basis method, but there
is a whole infrastructure for its implementation (like representation
of $u$-fractions, order among $x$-terms etc), so the remaining work is
relatively small.

OpenGeoProver consists of two main modules: one provides support for
algebraic methods and the other is a set of APIs from different
geometric applications and formats to the prover.
OpenGeoProver produces detailed reports with steps that are taken in
proving process: transformation of input geometric problem to algebraic
form, invoking a specified algebraic-based theorem prover, and presenting
the result with time and space spent to prove the theorem and with a
list of NDG conditions obtained during proving process, transformed to
a readable, geometry form.

One of the main purposes of this Java implementation is integration with
various dynamic geometry tools (including {\it GeoGebra}) that currently
don't have support for proving geometry theorems.
The architecture of OpenGeoProver enables easy integration with other
systems for interactive geometry and can be simply modified to accept
various input formats for conjectures.

%-----------------------------------------------------------------
\subsection{Representation of Geometry Conjectures and
Algebraization of Geometry Conjectures}
%-----------------------------------------------------------------

OpenGeoProver uses its internal structure to store information
about the given conjecture. This structure is called CCP
(Conjecture Construction Protocol) inspired by {\it GeoGebra}'s
Construction Protocol that contains the list of all constructions used
in building the scene in {\it GeoGebra}. Likewise, CCP contains the
list of all construction steps that build the scene for the conjecture
and, in addition, it contains the goal that has to be proved. Also,
CCP contains an algebraic form of the conjecture in the form of a
set of polynomials. In addition, CCP stores information about NDG
conditions for the conjecture, once they are returned from the
algebraic theorem proving method.

CCP is responsible for assigning symbolic coordinates to
constructed points and for the process of algebraization of geometry
conjectures. This processes are performed in the way outlined in
Section \ref{sec:Algebraization}: constructed points are instantiated
by pairs of symbolic variables that represent their coordinates, and
polynomials are generated from conditions that some point belongs to
some set of points (e.g.~intersection point of line and circle belongs
to that line and to that circle).

Although it would be sufficient to provide support for only a few
types of construction steps, for convenience CCP can deal with various
types of constructions and statements, used in various geometry
systems. Currently, it supports more than $20$ types of constructions
steps and $20$ types of statements.  Even if some construction isn't
currently supported, it may be possible to express it by two or more
existing construction steps. Also, it is possible to easily further
extend these sets.

%-----------------------------------------------------------------
\subsection{Algebraic Algorithms and Non-degeneracy Conditions}
%-----------------------------------------------------------------

Non-degeneracy conditions (in algebraic terms) are computed as
outlined in Section \ref{subsec:ndg}. There are algorithms that
transform algebraically specified conditions from polynomial form to
its geometry counterpart (\cite{BotanaValcarce,geother}). The idea for
our, currently implemented, algorithm for transformation of NDG
conditions is as follows:
\begin{itemize}
\item all variables that appear in polynomial form of a NDG condition
  are extracted;
\item for each of these variables, all points from CCP whose some
  coordinate has been instantiated with it are extracted;
\item combinations of these points are generated, but taking care to
  cover all variables;
\item these combinations of points are extended by points with zero
  coordinates (since they are neutral --- they could not introduce new
  variables to existing set of extracted variables);
\item for each combination, a number of possibilities is considered,
  depending on the number of points (e.g., for $3$ points $A$, $B$
  and $C$ we have the following specific positions: they are collinear,
  or $C$ is on circle with center $A$ and one point $B$, or line $AB$
  is perpendicular to $AC$, etc.);
\item for each specific possibility, its polynomial form is considered
  and checked whether it matches the input NDG polynomial.
\end{itemize}

Sometimes, it is possible to get several positions among points that satisfy
the same NDG polynomial, and we choose the one that includes free or random
points, i.e. the one that can be easily adjusted to affect other
constructed objects.

Note that the above procedure is not complete and there are some
NDG conditions that cannot be translated to the geometry form.

%-----------------------------------------------------------------
\subsection{Implementation Issues}
%-----------------------------------------------------------------

Algebraic primitives, like variables, powers, terms and polynomials,
are key elements for algebraic theorem provers and it is essential
to use data structures suitable for efficient algorithms. In OGP,
polynomial terms, for example, are organized as sorted lists of powers
of variables in descending order. This enables easy and efficient
comparison of terms and searching for a power. Polynomials are
organized as self-balanced binary search trees (i.e.,~objects of
the \verb|java.util.TreeMap| class) and this ensures efficient multiplying
of polynomials which is one of critical operations in Wu's method.

Geometry objects include geometry constructions, theorem statements
and NDG conditions. Each geometry construction and each statement
have their algebraic form which is used to transform the input
problem to polynomials in order to apply the prover. All geometric
constructions are stored in an indexed list and a construction map, all
within CCP which contains input
theorem. This enables simple access to a specific construction step
either by its index (or by the label of constructed object which must be
unique). NDG conditions initially have only polynomial form, and after
the prover's completed execution they are transformed to relationships
among geometry objects. To determine that relationship, OGP uses
objects for theorem statements, which already describe relationships,
and transforms them to the polynomial form to check if it matches the
polynomial for the specific NDG condition.

%-----------------------------------------------------------------
\subsection{Experimental Results}
%-----------------------------------------------------------------

There are several implementations of automated geometry theorem provers,
including provers based on algebraic methods but only the newest {\it Java
Geometry Expert} (JGEX) is open-source and self-contained (does not require
other systems or libraries). Because of that, we made a limited experimental
comparison between OGP and JGEX (using Wu's method), as the most closely
related tools (also both implemented in Java). Despite similarities, there
are also differences between these two systems. In contrary to OGP, JGEX
is developed for a specific geometry tool in mind that provides a graphical
interface to the underlying provers and additional features such as proof
visualization. Concerning the provers itself, OGP currently implements
only the partial Wu's method, while JGEX implements the complete Wu's method
(so, there are theorems that can be proved by JGEX but not by OGP).
In JGEX, geometric constructions are transformed to polynomial form and
triangulated one by one while objects are being constructed by the
user in the drawing panel. When the user press the button to prove a
theorem, only the final reminder has to be calculated.

For experimental comparison between JGEX and OGP, we selected, rather
arbitrarily, ten theorems in realm of both provers from Chou's {\em Collection}
\cite{chou}. We compared only the time of the final pseudo reminder calculation
(due to the JGEX style of operation), which is basically the most time-consuming
part of Wu's method. Results, given in Table \ref{table:exp}, of this limited
experimental comparison, of course cannot be conclusive, but they suggest that
OGP is comparable with JGEX (or even slightly more efficient).

\begin{table}
\begin{tabular}{|l|r|r|} \hline
Theorem                                                                                 & JGEX & OGP \\ \hline \hline
Example 6 (variant of; Pascal's thereom for circle)                                     & 1077 & 138 \\ \hline
Example 36 (butterfly)                                                                  & 1745 & 289 \\ \hline
Example 133 (variant of; orthocenter is incenter of triangle formed of altitudes' feet) &    7 &   7 \\ \hline
Example 154 (incenter)                                                                  &   38 &  15 \\ \hline
Example 173 (orthocenter)                                                               &    9 &   6 \\ \hline
Example 191 (Euler's line)                                                              &   10 & 162 \\ \hline
Example 196 (Euler's/nine-point circle)                                                 &    9 &  50 \\ \hline
Example 288 (Simson's theorem)                                                          &   25 &  16 \\ \hline
Example 336 (Gergonne's point)                                                          &   58 &  51 \\ \hline
Example 346 (Desargues' theorem)                                                        &   21 & 221 \\ \hline
\end{tabular}
\caption{Results of experimental comparison between OGP and JGEX (all execution
times are given in milliseconds}
\label{table:exp}
\end{table}

% ***************************************************************************
\section{Future Integrations}
\label{sec:integrations}
% ***************************************************************************

\begin{figure}[ht!]
\begin{center}
\input{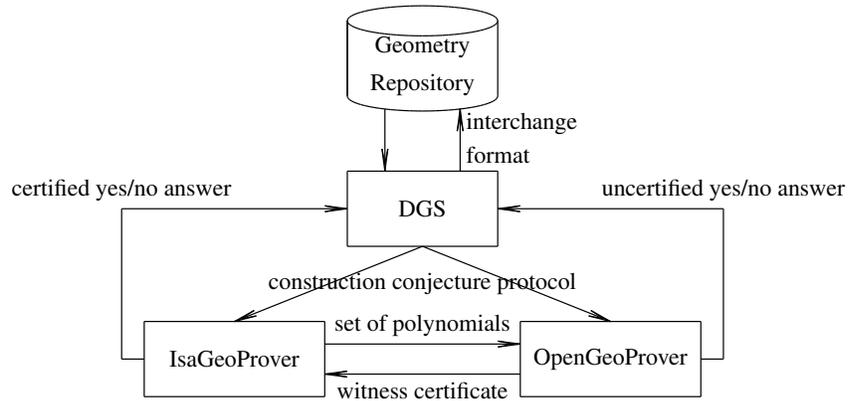}
\end{center}
\caption{Integration of DGS, Isabelle geometry prover and OpenGeoProver}
\label{fig:big_picture}
\end{figure}

The components described above could and should be integrated with
other tools and resources, as illustrated in Figure
\ref{fig:big_picture}.  A dynamic geometry tool can enable user to
enter geometry constructions and conjectures or to import them from
some repository (e.g.,
GeoThms\footnote{\url{http://hilbert.mat.uc.pt/~geothms/}}).  The user
should also be able to export his/her construction to the repository,
and this communication should be performed using some of the emerging
standard interchange formats (e.g., i2g). The dynamic geometry tool
can invoke an automated theorem prover on demand (to check if some
conjecture is indeed a theorem), or to check if some construction is
regular \cite{regular-constructions}. In either case, a conjecture, in
the form of conjecture construction protocol, is sent to Isabelle (if
a verified answer is required) or to OpenGeoProver (if efficiency is
more important than a verified answer). If invoked, our geometry
prover in Isabelle calls OpenGeoProver (or some other algebraic tool)
and asks for a witness certificate (e.g., a Gr\"obner basis) that can
be used for verified proving of the given conjecture. Finally, the
yes/no answer and NDGs are returned by Isabelle prover or
OpenGeoProver back to DGS. This interaction can bring new features to
existing DGS and other related systems. For instance, compared to
{\it GeoProof}, the proposed system would also provide efficient Java
provers, while, compared to {\it GCLC}, the proposed system would
provide link to a proof assistant and verified reasoning.

The proposed system with described interaction can be relevant for new
applications in education. However, its components can be also useful
in formal explorations of geometry. Namely, the algebraic methods
implemented as Isabelle tactics would be helpful for many intermediate
tasks along building a formalized body of geometry knowledge.

% ***************************************************************************
\section{Applications in Education}
\label{sec:education}
% ***************************************************************************

Currently, automated and interactive theorem proving is used in education
(especially, in pre-university education) only very sporadically and in a
very limited scope. Despite that fact that theorem proving can be very
demanding for pupils/students to deal with (to use and understand), we
believe that theorem proving can be used in a realistic and beneficial way.
In the following we list some potentials for applications of the presented
integrated framework, outlined in Section \ref{sec:integrations}, in
education.

\begin{itemize}
\item Geometry has been in education widely used as a theory suitable
for introducing the notions of axiom system and proof. Students are
typically understand and exercise logical inferences on a concrete
theories (e.g., Euclidean geometry) and on concrete examples. The
same could and should be applied for introducing the concept of
{\em formal, machine verifiable proof} to the students. Students
of the modern, computer era should understand that proofs are strict,
formal objects subject to verification by a computer, especially
bearing in mind expectations that in some decades a significant
portion of mathematics will be machine verifiable. In this context,
as with ``informal proofs'', geometry is very suitable (because
of its intuitive and visual nature). Having a setting in which only
some relevant fragments have to be provided by the student (and
the rest provided by already developed infrastructure), make this
learning scenario realistic and useful, of course, if suitable
adapted to specific cases.

\item In mathematical education, mathematical knowledge requires
{\em justification} and {\em explanation} for conjectures. Algebraic
methods in geometry, however, provide just the former one. Still,
they (within a proof assistant or as a stand-alone implementations)
can be used to help students for filling gaps in their proofs.
This way, some gaps can turned to be based on wrong conjectures,
or may reveal additional conditions needed for the conjecture
to be valid.

\item With most of dynamic geometry tools, a student typically
explores his/her construction, trying to visually or numerically
detect or check that certain conjecture is always true. However,
there is no mathematical argument in such conclusion and this
may even lead to some misconceptions in student's understanding
of geometry. Instead of this ``visual checking'', a theorem prover
can be used for bringing precise, mathematical conclusions.
This way, a DGS can still keep its role, but now enriched
with strict mathematical arguments. In this scenario, either
an algebraic method formalized within a proof assistant or
(preferably) a light-weight, stand-alone automated theorem
prover integrated within the DGS can be used.

\item In mathematical education, geometry construction problems
are one of the standard (and, in the same time) most challenging
types of problems: given some constraints, the student is asked
to construct a geometry figure meeting these constraints.
The construction is often accompanied by some kind of analysis,
visualization and an informal proof. Still, these proofs often
tend to be flawed, most often due to degenerate cases. In this
scenario, the student could used an algebraic method for checking
his/her construction, before even attempting to prove that it is
correct. If it is, the student have to prove it in a way that
is expected from him/her.
\end{itemize}

We stress that our presented framework currently does not provide
an interface needed for applications in education, but give all
key infrastructure.

% ***************************************************************************
\section{Conclusions}
\label{sec:conclusions}
% ***************************************************************************

In this paper we presented our ongoing project on implementation and
formalization of algebraic methods for geometry theorem proving.  The
project should hopefully help wider applications of the algebraic
methods in education and formalization of mathematics.

%for example:
%http://dx.doi.org/10.1017/CBO9780511565847
\providecommand{\urlalt}[2]{\href{#1}{#2}}
\providecommand{\doi}[1]{doi:\urlalt{http://dx.doi.org/#1}{#1}}

\end{document}